\documentclass[aps, prl, nofootinbib, amsmath, twocolumn, preprintnumbers, supserscriptaddress]{revtex4-1}
\usepackage{amssymb,esvect,amsmath,graphicx,latexsym,amsthm,slashed,eso-pic,hyperref,bm, url}
\usepackage[shortlabels]{enumitem}
  \DeclareMathOperator{\mev}{MeV} \DeclareMathOperator{\gev}{GeV}       \DeclareMathOperator{\s}{s}     \DeclareMathOperator{\erg}{erg}  
 \newcommand{\cC}{{\cal C}} \newcommand{\cD}{{\cal D}}  \newcommand{\cF}{{ \cal F}}     \newcommand{\cM}{{\cal M}}     \newcommand{\cS}{{\cal S}}

\newcommand{\pL}{\left(} \newcommand{\pR}{\right)} \newcommand{\bL}{\left[} \newcommand{\bR}{\right]}    
\newcommand{\beq}{\begin{equation}} \newcommand{\eeq}{\end{equation}}
\newcommand{\bea}{\begin{eqnarray}} \newcommand{\eea}{\end{eqnarray}}

\newcommand{\tenx}[1]{\times 10^{#1}}
\newcommand{\Eq}[1]{Eq.~(\ref{#1})}  
  
\newcommand{\Fig}[1]{Fig.~\ref{#1}} 
\newcommand{\Tab}[1]{Tab.~\ref{#1}}

\begin{document}

\preprint{FERMILAB-PUB-19-575-T}

\title{A New Mask for An Old Suspect:\\Testing the Sensitivity of the Galactic Center Excess to the Point Source Mask}
\author{Yi-Ming Zhong${}^a$}
\author{Samuel D.~McDermott${}^b$}
\author{Ilias Cholis${}^c$}
\author{Patrick J.~Fox${}^b$}
\affiliation{${}^a$Kavli Institute for Cosmological Physics, University of Chicago, Chicago, IL
, USA}
\affiliation{${}^b$Fermi National Accelerator Laboratory, Batavia, IL, USA}
\affiliation{${}^c$Department of Physics, Oakland University, Rochester, MI
, USA}

\begin{abstract}
The Fermi-LAT collaboration has recently released a new point source catalog, referred to as 4FGL. For the first time, we perform a template fit using information from this new catalog and find that the Galactic center excess is still present. On the other hand, we find that a wavelet-based search for point sources is highly sensitive to the use of the 4FGL catalog: no excess of bright regions on small angular scales is apparent when we mask out 4FGL point sources. We postulate that the 4FGL catalog contains the large majority of bright point sources that have previously been suggested to account for the excess in gamma rays detected at the Galactic center in Fermi-LAT data. Furthermore, after identifying which bright sources have no known counterpart, we place constraints on the luminosity function necessary for point sources to explain the smooth emission seen in the template fit.
\end{abstract}

\maketitle

\noindent \textit{\textbf{Introduction:}}
An excess of gamma rays has been detected near the Galactic Center \cite{Goodenough:2009gk, Hooper:2010mq, Hooper:2011ti} (the ``GCE''), extending to Galactic latitudes greater than $10^\circ$ above the midplane \cite{Hooper:2013rwa, Daylan:2014rsa}. This excess appears to be robust to uncertainties in the modeling of expected diffuse emission \cite{Abazajian:2014fta, Calore:2014xka}, yet its origin remains debated. While originally interpreted as evidence of dark matter annihilation \cite{Goodenough:2009gk, Hooper:2010mq, Daylan:2014rsa}, the excess may have features suggestive of an origin in a population of point sources \cite{Hooper:2010mq, Abazajian:2010zy, Lee:2014mza, Lee:2015fea, Bartels:2015aea, Chang:2019ars}. In particular, Ref.~\cite{Bartels:2015aea} proposed a candidate population of sufficiently luminous point sources using wavelet-based techniques.

Recently, Ref.~\cite{Leane:2019xiy} has found evidence that a mismodeled population of point sources could act as a source of bias in the techniques of Refs.~\cite{Lee:2014mza, Lee:2015fea}. Nonetheless, the results of Ref.~\cite{Bartels:2015aea} appear less sensitive to the potential for bias in statistical methods, so one may reasonably conclude that such bias is not after all a concern for the statistical methods employed thus far.

In this paper, we provide evidence that the large majority of the point sources originally found in \cite{Bartels:2015aea} have subsequently been independently discovered and characterized by the Fermi-LAT collaboration as members of the 4FGL point source catalog \cite{Fermi-LAT:2019yla}: in some sense, \cite{Bartels:2015aea} predicted parts of the 4FGL catalog. Given the size and quality of the 4FGL catalog, and given the claim that the point sources identified in \cite{Bartels:2015aea} were members of a population bright enough to account for the GCE, it is important to reevaluate evidence for the GCE using a mask based on 4FGL sources. The 4FGL catalog provides spectral information for these point sources, so an updated template-based search for the GCE across a wide energy range is now warranted. 

In this paper, we first perform a template fit to Fermi data and then search for small-scale power therein. We find that the GCE remains preferred at high statistical significance in the template fit, and, furthermore, the normalization of the GCE does not appreciably decrease, beyond small finite-area effects, when we move from a 2FGL to 4FGL mask. However, the amount of small-scale power decreases almost entirely. Because the ``small-scale excess'' goes away when including a mask of 4FGL sources but the GCE is almost unchanged, we argue that this is evidence that the GCE is not due to bright point sources.

\begin{figure*}[t]
\begin{center}
\includegraphics[width=0.475\textwidth]{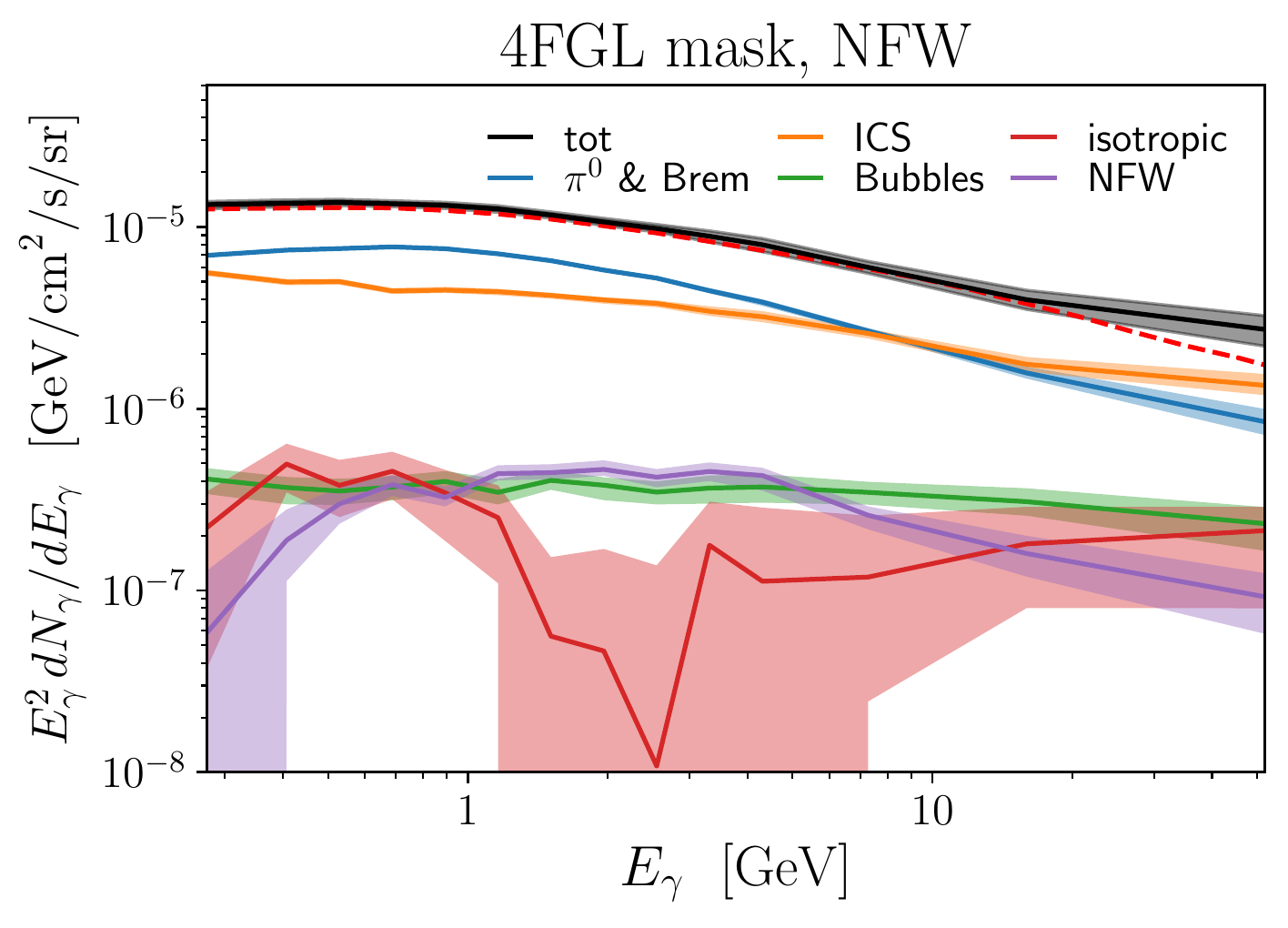}~~~
\includegraphics[width=0.475\textwidth]{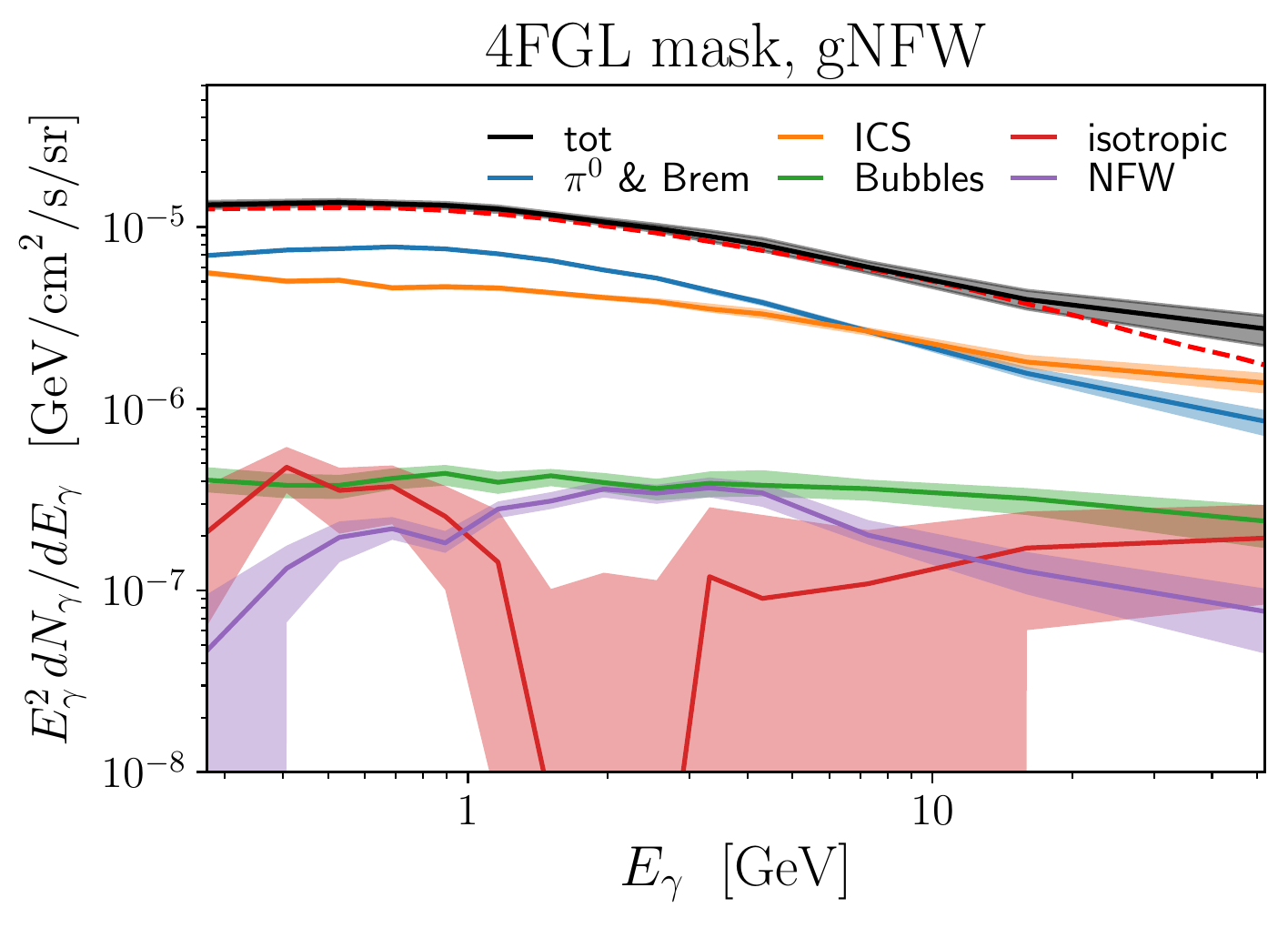}
\caption{Results of a template fit including a GCE with an NFW profile {\bf (left)} and a gNFW profile {\bf (right)}, masking the entire 4FGL catalog. Shaded bands show approximate $95\%$ preferred regions from our MCMC analysis. The dashed red line is the expected total diffuse model. The GCE, with either morphology, is incompatible with zero at high statistical significance.}
\label{template_fit_result}
\end{center}
\end{figure*}

~\\[-\baselineskip]
\noindent\textit{\textbf{Template Fit With 4FGL:}}
Motivated by the recent introduction of the 4FGL point source catalog \cite{Fermi-LAT:2019yla}, we perform a template analysis of gamma-ray data (up to week 559 of the Fermi-LAT mission) after masking all 4FGL sources. For this analysis, we restrict to a single diffuse template model, Model A from \cite{Calore:2014xka}, which was shown to reproduce the fitted emission from the galactic center with small deviations from the predicted diffuse component fluxes. Model A contains contributions from inverse Compton scattering, $\pi^0$, bremsstrahlung, the Fermi bubbles, and an isotropic component.  We fix the relative normalizations of $\pi^0$ and bremsstrahlung components. We leave for future study a detailed analysis of the sensitivity of the extracted normalizations to background models.

 On top of Model A, we test two models of excess emission: either a smooth NFW profile with a dependence on galactocentric distance of $\rho_{\rm NFW}^2$, where $\rho_{\rm NFW} \propto (r/r_s)^{-1}[1+r/r_s]^{-2}$; or a smooth generalized NFW profile with dependence on galactocentric distance of $\rho_{\rm gNFW}^2$, where $\rho_{\rm gNFW} \propto (r/r_s)^{-\gamma}[1+r/r_s]^{-3+\gamma}$ and inner slope $\gamma = 1.2$. We make two choices of point source mask: 
the 2FGL mask (which was current at the time of \cite{Daylan:2014rsa, Abazajian:2014fta, Calore:2014xka}) and the 4FGL mask.
With these templates, we use the {\tt emcee} MCMC program \cite{ForemanMackey:2012ig}, to maximize the likelihood $\lambda$ that the data is described by our model with parameters $\bm  \theta$.  We consider $0.1^\circ\times0.1^\circ$ pixels in the region with galactic longitude $|\ell| \leq 20^\circ$ and latitude $2^\circ \leq |b| \leq 20^\circ$. The negative log likelihood for Poisson-distributed data is \cite{Tanabashi:2018oca}
$-2\ln \lambda(  \bm  \theta)
= 2 
\sum_{i \in {\rm pixels}} \! \bL \mu_i \big( \bm  \theta \big) - n_i + n_i \ln \frac{n_i}{\mu_i( \bm \theta )}\bR + \chi^2_{\textrm{ext}},$
where the sum goes over the $360\times400$ pixels in our region of interest, for each of our 14 energy bins. Due to the drop in statistics at higher photon energies, we use 11 evenly log-spaced energy bins covering observed energies between $0.275$ GeV and 4.91 GeV with 3 additional wider energy bins from 4.91 GeV to 51.9 GeV.   We fit each energy bin independently.  In 
$-2\ln \lambda(  \bm  \theta)$, the term $\chi^2_{\textrm{ext}}$ includes external constraints on the spectral properties of the Fermi Bubbles and isotropic gamma-ray spectra as in \cite{Calore:2014xka}.

The Fermi PSF varies with energy so the masks we use also vary.  At each energy bin we use a mask with a TS-dependent size, similar to \cite{Bartels:2015aea} as follows.  We place a smaller mask whose radius decreases monotonically from $\sim 1^\circ$ at the lowest energy to half of our pixel size ($0.05^\circ$) at the highest energy at each source in the 4FGL point source catalog with a Fermi-LAT TS between 9 and 49.  We place an approximately 3 times larger mask for sources with a Fermi-LAT TS greater than 49.
 
Results for our NFW and gNFW analyses with 4FGL mask are displayed in \Fig{template_fit_result}. The units of these curves are normalized against the area of our $|\ell| \leq 20^\circ$, $2^\circ\leq|b|\leq 20^\circ$ region of interest, of size $0.43$ sr. We compare our best fit spectrum to the {\tt GALPROP} predicted total energy spectrum (dashed red line). The shaded purple region, representing the GCE with spatial dependence $\rho_{\rm NFW}^2$ or $\rho_{\rm gNFW}^2$, differs from zero at high significance. The normalization of this new component depends at the $\sim 30\%$ level on the spatial profile, but the preference for a nonzero GCE is statistically very significant in both cases. We provide values of $-2 \ln \lambda$ (summed over energy bins) in \Tab{ll-values}, using the best fit values of the component normalizations $\bm \theta$ for each model and each energy bin, as determined by our MCMC analysis.

\begin{table}[b]
\caption{Difference in $-2\ln\lambda$ (lower numbers are better) at the best fit points of each model, summed over energy bins, compared to our best fit for each mask.}
\begin{center}
\begin{tabular}{c|c|c|c}
Type of Mask & NFW & gNFW & no excess \\ \hline
2FGL & - & 476 & 5430 \\
4FGL & - & $368$ & $3600$
\end{tabular}
\end{center}
\label{ll-values}
\end{table}%

~\\[-\baselineskip]
\noindent\textit{\textbf{Searching for Small-Scale Power:}}
Although the change in the mask from the 2FGL to the 4FGL catalog does not affect the preference for a new smooth emission component in our template fit, it is still interesting to ask how the new 4FGL mask impacts other searches at the Galactic center; {\it e.g.}, point source searches.

\begin{figure*}[t]
\begin{center}
\includegraphics[width=0.995\textwidth]{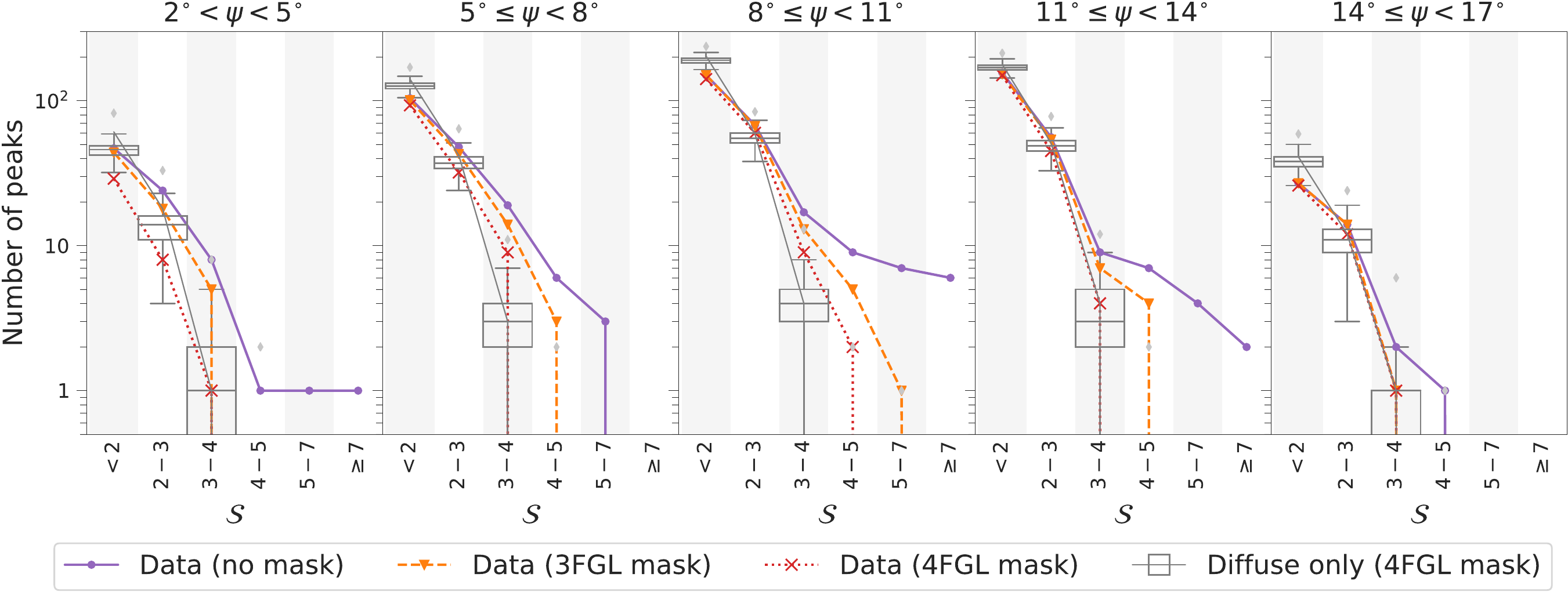}
\caption{Impact of different point source masks on our test statistic $\cS[\cD]$ defined in \Eq{SNR-M2} for $1 \gev \leq E_\gamma \leq 4 \gev$. We show results from analyzing data with: no mask (solid purple line), a mask of the 3FGL point source catalog (dash-dotted orange line), and a mask of the 4FGL point source catalog (dotted red line). We compare to $\cS[\cM_i]$ (gray box and whiskers). Because there are no pixels with $\cS[\cD] \geq 4$ when the entire 4FGL point source catalog is used (aside from two point sources at $8^\circ \leq \psi < 11^\circ$), all small-scale power must be documented in the 4FGL catalog.}
\label{mask-impact}
\end{center}
\end{figure*}

\begin{figure}[t]
\begin{center}
\includegraphics[width=0.485\textwidth]{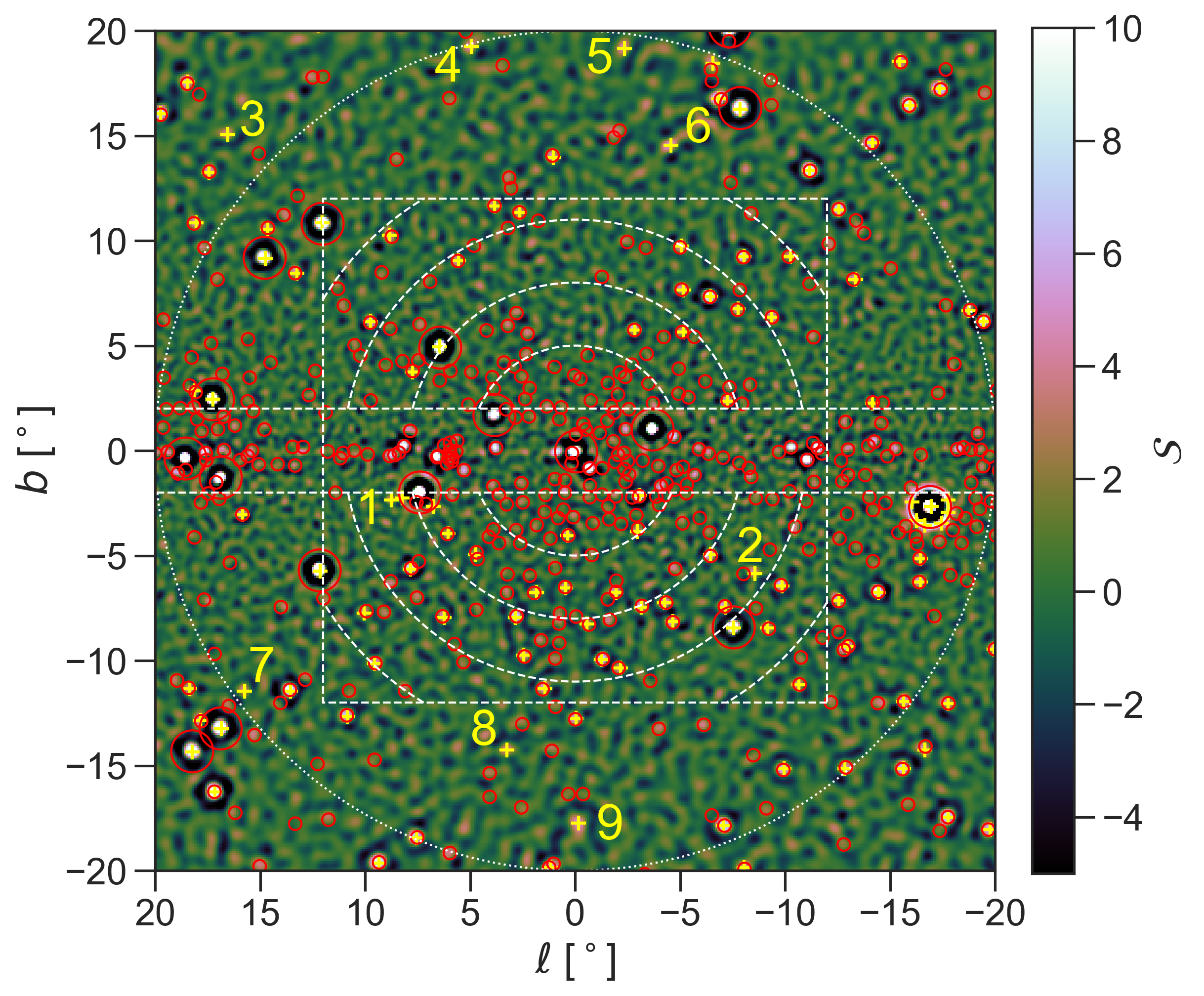}
\caption{Map of $\cS[\cD]$. Peaks with $\cS[\cD]\geq4$ in the region $2^\circ \leq |b| \leq 20^\circ$ and $| \ell |\leq 20^\circ$ are shown as crosses. 4FGL sources are shown in red circles of size $0.3^\circ (1^\circ)$ for $9\leq TS \leq49 \, (49\leq TS)$. We find 8 sources (numbers 1 and $3-9$) that have no noticeable 4FGL counterpart, and one (number 2) that is sensitive to association proximity cut. The inner dashed white lines show the angular regions used in Figs.~\ref{mask-impact}, \ref{compare-Bartels}, \ref{dm-impact}, and \ref{csp-impact}; the outer dashed white lines shows the maximum extent of a region of projected galactocentric distance 3 kpc. See text for details.}
\label{S4-locations}
\end{center}
\end{figure}

We use a matched-filter algorithm based on a Mexican hat function, which we refer to as a wavelet, to identify point sources in the region described by $2^\circ \leq|b| \leq 12^\circ$ and $|\ell| \leq 12^\circ$ and the energy range $1 \gev \leq E_\gamma\leq4\gev$, as in \cite{Bartels:2015aea}.  
We evaluate the test statistic defined by \cite{Bartels:2015aea} on a map of photon counts and bin the resulting maps by values of $\cS$ in each pixel. (We reproduce the definitions of \cite{Bartels:2015aea} in our appendix; the test statistic $\cS$ is defined in \Eq{SNR-M2}.) In \Fig{mask-impact} we compare the results of this procedure using a map of actual photon counts $\cD$ and on many Poisson samples of expected counts from a range of diffuse models $\cM_i$. The different annuli use, each of width $3^\circ$ and within, $2^\circ\leq |b| \leq 12^\circ$ and $|\ell| \leq 12^\circ$, are shown in \Fig{S4-locations}.

For $\cM_i$, we use 60 models from \cite{Calore:2014xka}, produced in $0.25^{\circ} \times 0.25^{\circ}$ bins on a Cartesian grid. They were generated using {\tt GALPROP} v54 WebRun \cite{VLADIMIROV20111156, GALPROPSite}, re-binning into $0.1^{\circ} \times 0.1^{\circ}$ bins. We first calculate counts and expected diffuse emission fluxes, then smooth these model maps we use the Fermi-LAT PSF profile information for FRONT data \cite{FermiPSF}. We also include  the isotropic gamma-ray emission from \cite{Ackermann:2014usa} and the Fermi Bubbles emission from \cite{Su:2010qj} (morphological information) and \cite{Fermi-LAT:2014sfa} (spectral information). We Poisson sample each diffuse model 100 times, allowing the diffuse component normalizations to vary within $15\%$ of the total observed gamma-ray emission within $2^\circ\leq |b| \leq 20^\circ$ and $|\ell| \leq 20^\circ$. We show the results of the binned $\cS$ maps of our diffuse models as box-and-whiskers plots. These plots display features of the distribution of the $60\times100$ mocks that we generate: the central line is the median value of this distribution, the box shows the interquartile range of this distribution (spanning 25\% to 75\% containment), and the whiskers show 1.5 times the (asymmetric) interquartile range. The most extreme outlier from each set of samples is shown as a faint point if it lies outside of the whiskers. One qualitative trend that is clear from the diffuse-only analysis is that the diffuse-only models are able to produce pixels with $\cS$ values up to 4, but only extreme outliers produce any pixels with $\cS[\cM_i] \geq 4$.

We compare these model expectations to a map of $\cS$ derived from data. First we do this with no point source mask.
From \Fig{mask-impact}, we see that the data contains a large number of pixels with $\cS[\cD] \geq 3$ compared to the diffuse-only expectation. It also copiously produces $\cS[\cD] \geq 4$ pixels, which diffuse-only models cannot do. Of course, many of the pixels with $\cS[\cD] \geq 4$ host known point sources. In some cases, these point sources are identified in multiple wavelengths, and are known to have characteristics that make them unsuitable candidates to explain the GCE. Thus, we should partially mask our map of $\cS[\cD]$.

To study the impact of masking, we alternately choose two masks for the data. First, we consider the 3FGL point source catalog \cite{Acero:2015hja}, with a TS-dependent size, as discussed earlier.  Following \cite{Bartels:2015aea}, we place a mask of $0.3^\circ$ at each source in the 3FGL point source catalog with a Fermi-LAT TS between 9 and 49, and a mask of $1^\circ$ at each source in the 3FGL point source catalog with a Fermi-LAT TS greater than 49. Next, we consider a mask using the entire 4FGL \cite{Fermi-LAT:2019yla} point source catalog, with the same sizing convention.  
In the appendix, we also compare to a different mask advocated for in \cite{Bartels:2015aea}, the results of which we are able to reproduce.

As we show in \Fig{mask-impact}, upon applying a mask for all 4FGL point sources (dotted line), there are only two pixels with $\cS[\cD] \geq 4$ in the $2^\circ\leq |b| \leq 12^\circ$ and $|\ell| \leq 12^\circ$ region of interest. Because of the small number of surviving high-$\cS[\cD]$ peaks when the entire 4FGL point source catalog is used, we conclude that {\it all bright point sources in the Fermi-LAT data} (aside from the two new sources) {\it have a counterpart in the 4FGL catalog}. The 4FGL mask makes a qualitative difference in this search for small-scale power, despite the fact that it does not impact the preference for a GCE in the template fit, as shown in \Tab{ll-values}.  For energies $\simeq 1.5\gev$, the 4FGL mask blocks about 3 times more of the inner sky than the 2FGL mask, while beyond $11^\circ$ it is only a factor of 2 more.  From the inner to outer regions the 4FGL mask covers $47\%, 24\%, 16\%, 12\%, 13\%$ of the sky, which is unlikely to explain the drop in $\cS>4$ pixels by accident alone.

~\\[-\baselineskip]
\noindent\textit{\textbf{Possible Interpretations:}}
Because the wavelet-based search we implemented above is good at finding isolated bright pixels, the results presented so far are compatible with an underlying smooth distribution of excess photons; we demonstrate this explicitly in the representative case of dark matter annihilation in the appendix. Here, we use our results to uncover some information about the nature of a putative central source population (CSP) which would account for the excess \cite{Bartels:2015aea}.

Because we are able to localize the high-$\cS[\cD]$ pixels at the sub-$0.1^\circ$ level, we begin by searching for counterparts to the $\cS[\cD]\geq4$ pixels in the $|\ell|\leq20^\circ, 2^\circ \leq |b|\leq20^\circ$ box that we used for our template analysis. We show the map of $\cS[\cD]$ in \Fig{S4-locations}. We find $115$ pixels in this region with $\cS[\cD] \geq 4$. We find that 107 of these are near a 4FGL source; we highlight the 9 points with $\cS[\cD] \geq 4$ that are more than $0.3^\circ$ away from a 4FGL source with numbers in \Fig{S4-locations}, and thus are unmasked in the analysis of \Fig{mask-impact}. The source labelled 2 is the closest of these; to be conservative we connect it to the unassociated source 4FGL J1750.0-3849, which is $0.53^\circ$ distant. All other sources are at least $0.7^\circ$ from a $\cS[\cD] \geq 4$ pixel.

Of these pixels with counterparts in the 4FGL catalog, $\sim 40$ of the counterparts are unknown or unassociated sources which we consider CSP candidate members, or are Galactic center millisecond pulsars, which are potential CSP members. The uncertainty on the number of counterparts comes from the fact that this counting is sensitive to our procedure for making associations. Then number varies from 37 to 47 for different reasonable choices for proximity cut, which are described in the appendix. We compare the GCE from \Fig{template_fit_result} to the stacked spectra of $47$ 4FGL sources. For the stacked spectrum, we include a $20\%$ error bar, which approximately accounts for: uncertainties in the magnitude of the 4FGL sources, which are $\leq 12\%$ for the bright sources; the absence of spectra for the 8 $\cS[\cD]\geq4$ pixels with no nearby 4FGL source, some of which may be suitable for inclusion in this analysis; and possible infelicities in our peak-finding and clustering algorithms, described in the appendix.

We see from \Fig{spectra-compare} that the peak of the GCE is a factor of $\sim 4$ higher than the stacked spectra of possibly interesting 4FGL sources, or that the integral of $E_\gamma dN_\gamma/dE_\gamma $ above  $687 \mev$ is a factor of 4.5 larger. Including a $\simeq 20\%$ error, we deduce that if the GCE originates in a CSP this CSP must have sub-threshold sources, which remain unmasked, that outshine the above-threshold sources, which are stacked and shown in \Fig{spectra-compare}, by a factor of $4\pm 1$. For a power law luminosity function of the form $dN/dL \propto L^{-\alpha_L}$ with hard cutoffs at $L_{\rm min},L_{\rm max}$, the ratio of total flux below and above the point source detection threshold is $\big[\big(\frac{L_{\rm thr}}{L_{\rm max}}\big)^{2-\alpha_L}-\big(\frac{L_{\rm min}}{L_{\rm max}}\big)^{2-\alpha_L}\big]/\big[1-\big(\frac{L_{\rm thr}}{L_{\rm max}}\big)^{2-\alpha_L}\big]$. If we take the upper cutoff to be exponential instead, the ratio of total flux below and above the point source threshold is $\big[\Gamma\big(2-\alpha_L, \frac{L_{\rm min}}{L_{\rm max}}\big) - \Gamma\big(2-\alpha_L, \frac{L_{\rm thr}}{L_{\rm max}}\big)\big]/\Gamma(2-\alpha_L,\frac{L_{\rm thr}}{L_{\rm max}}) $. Using fiducial values $L_{\rm thr}= 10^{34}\erg\!/\!\s, L_{\rm max} = 10^{35}\erg\!/\!\s$, we find $\alpha_L \gtrsim 1.88\pm0.03 \, (1.90\pm 0.02) $ for the exponential (hard) cutoff in the limit $L_{\rm min} \to 0$. For such a sharply falling power law, the physics that sets the lower cutoff becomes important, however. Imposing $L_{\rm min}=10^{29}\erg\!/\!\s$, we find $\alpha_L \gtrsim 1.93\pm0.04 \, (1.96\pm0.04)$ for the exponential (hard) cutoff. Changing the value of $L_{\rm thr}$ to $3\tenx{34}\erg\!/\!\s$ while taking the limit $L_{\rm min} \to 0$ and keeping $L_{\rm max}$ fixed, which is a very extreme set of choices, makes $\alpha_L \geq 1.78\pm0.05 (1.81\pm 0.05)$ for the exponential (hard) cutoff. These observations therefore rule out the range of $1.2 \leq \alpha_L \leq 1.5$ preferred by local observations of millisecond pulsars \cite{Cholis:2014noa, Petrovic:2014xra} and the value of $\alpha_L =1.5$ assumed in \cite{Bartels:2015aea}. After calculating $\alpha_L$ as above, one may calculate the corresponding number of sub-threshold sources required to produce the GCE. With $L_{\rm min}=10^{29}\erg\!/\!\s, L_{\rm thr}= 10^{34}\erg\!/\!\s, L_{\rm max} = 10^{35}\erg\!/\!\s$, we find $N_{\rm sub} = (3.1 \pm 1.3)\tenx6 \, [(3.6 \pm 1.5)\tenx6]$ for the exponential [hard] cutoff, compared to 47 sources above threshold.

\begin{figure}[t]
\begin{center}
\includegraphics[width=0.485\textwidth]{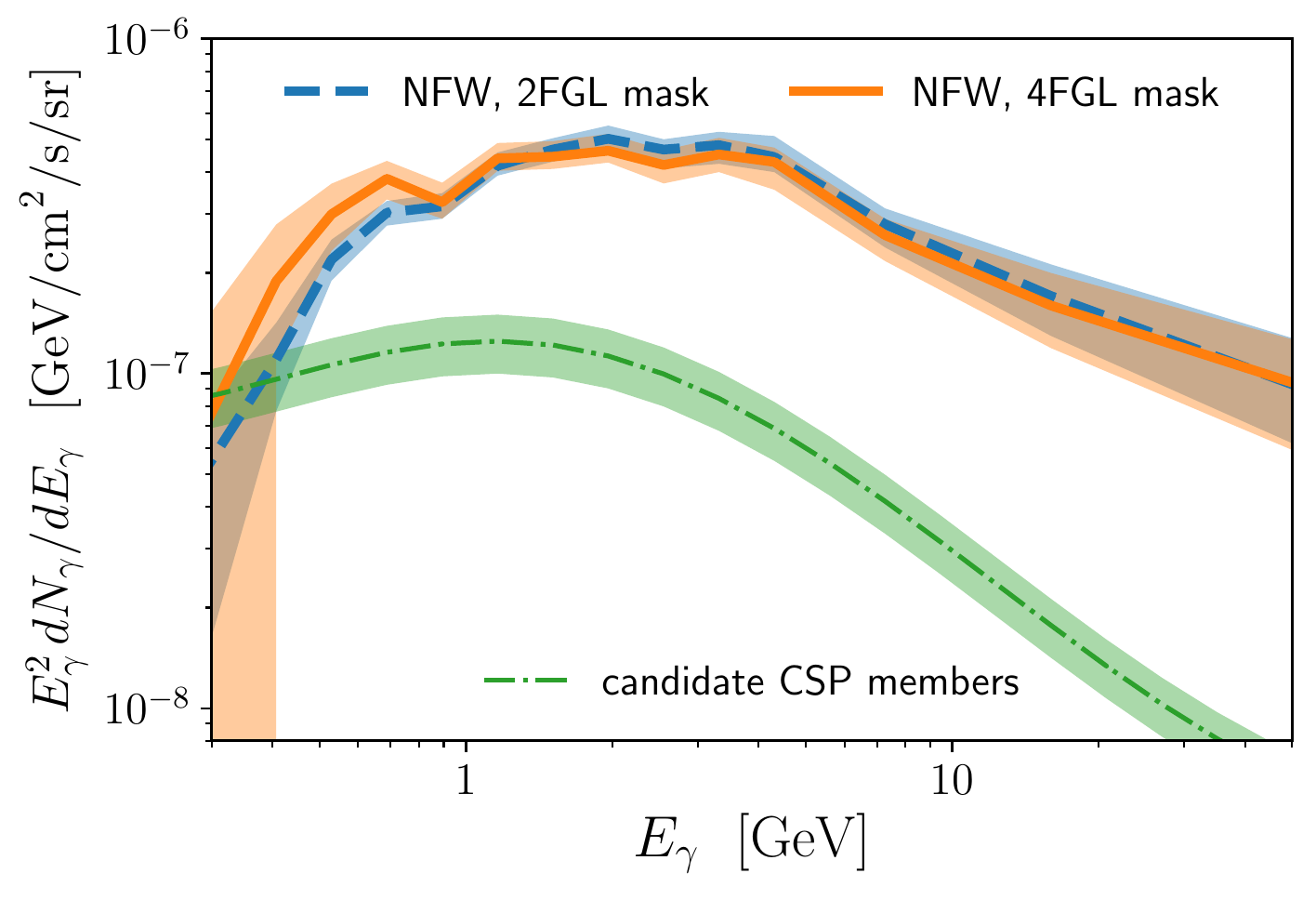}
\caption{Our best fit GCE with 2FGL mask (blue dashed) or 
4FGL mask (orange solid), compared with the stacked spectra of 4FGL sources that appear in our matched-filter search with $\cS > 4$ which are possibly members of a CSP (green dot-dashed).
}
\label{spectra-compare}
\end{center}
\end{figure}

~\\[-\baselineskip]
\noindent\textit{\textbf{Conclusions:}}
We have explored the impact of the 4FGL catalog on the GCE and on the proposed point source explanation thereof. We find that the GCE is still preferred at high statistical significance over no excess even when the 4FGL catalog is masked. However, our search for small-scale power using a wavelet-based test statistic is extremely sensitive to the presence of the 4FGL catalog: when 4FGL is masked, only two pixels with small-scale power remain in $2^\circ \leq |b| \leq 12^\circ$ and $| \ell |\leq 12^\circ$ region of interest. We demonstrated that this discordance between template- and wavelet-based searches implies that the GCE does not primarily originate in bright point sources, and for it to arise from a sub-threshold population of dim sources there is a strong constraint on the slope of the luminosity function.

Along the way, we have identified several high significance points with no presently known 4FGL counterpart, which may be uncovered in future point source catalogs. Further study of these sources, and further characterization of the GCE itself, is warranted in light of our results. The future of the study of the GCE appears to us quite bright.

\section*{Acknowledgments}
We thank Simona Murgia, Tim Tait, Jesse Thaler, and Christoph Weniger for helpful and friendly discussions. We thank Tim Linden, Mariangela Lisanti, Nick Rodd, Jesse Thaler, and Christoph Weniger for comments on a draft. SDM especially thanks Rebecca Leane, Tracy Slatyer, and Jesse Thaler for conversations and the CTP at MIT for hospitality while part of this work was conducted. PJF, SDM, and YZ thank the Aspen Center for Physics (which is supported by National Science Foundation grant PHY-1607611, and where this work was performed in part) for hospitality. PJF and SDM would like to express a special thanks to the GGI Institute for Theoretical Physics for its hospitality and support. PJF's work there was supported by a grant from the Simons Foundation (SIMONS, 341344 AL). Fermilab is operated by Fermi Research Alliance, LLC under Contract No. De-AC02-07CH11359 with the United States Department of Energy. This work was supported in part by the Kavli Institute for Cosmological Physics at the University of Chicago through an endowment from the Kavli Foundation and its founder Fred Kavli.

\appendix

\setcounter{equation}{0}
\renewcommand{\theequation}{A.\arabic{equation}}

\section{Data Selection}

\begin{figure*}[t]
\begin{center}
\includegraphics[height=0.28\textheight]{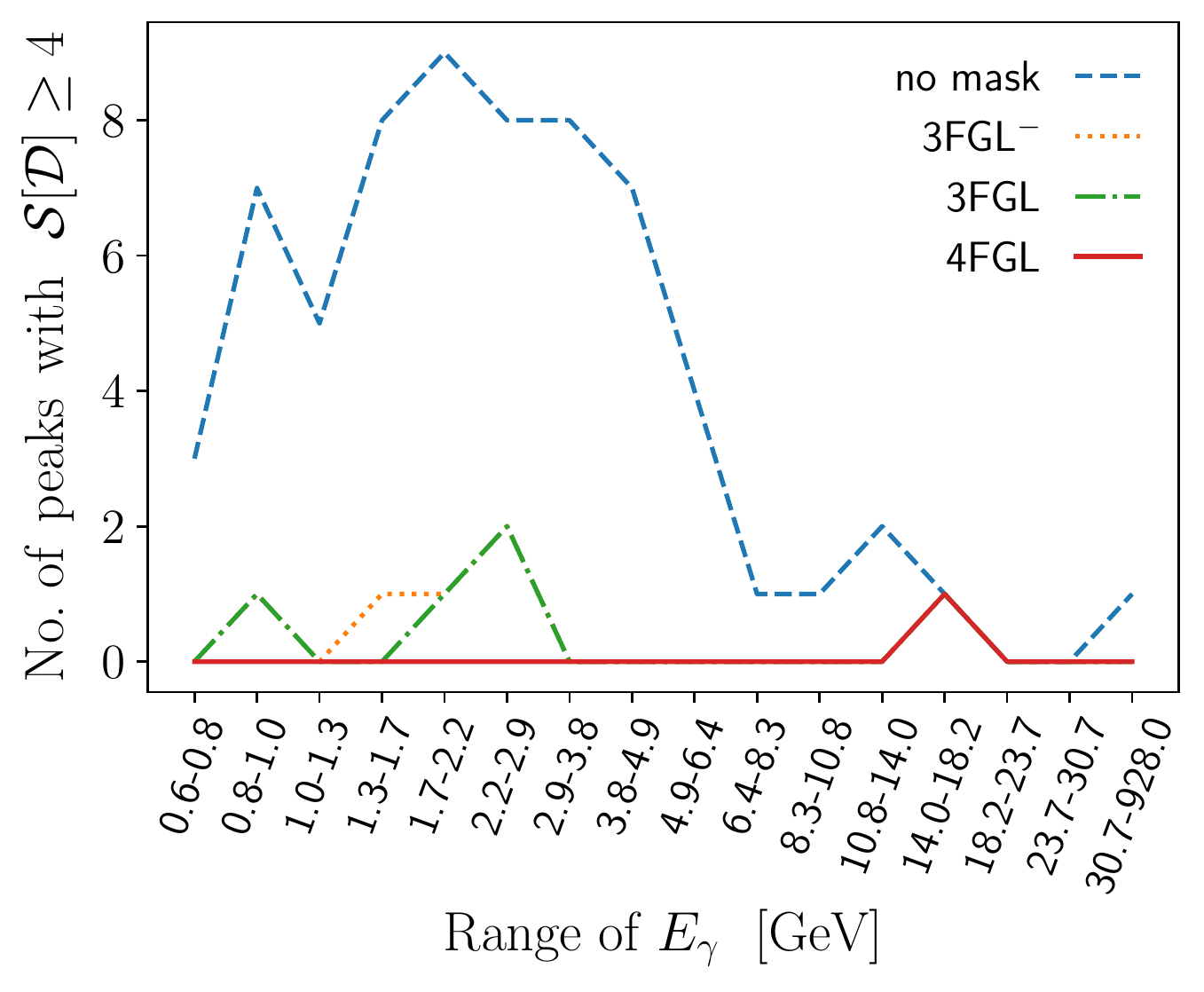}~~~~
\includegraphics[height=0.28\textheight]{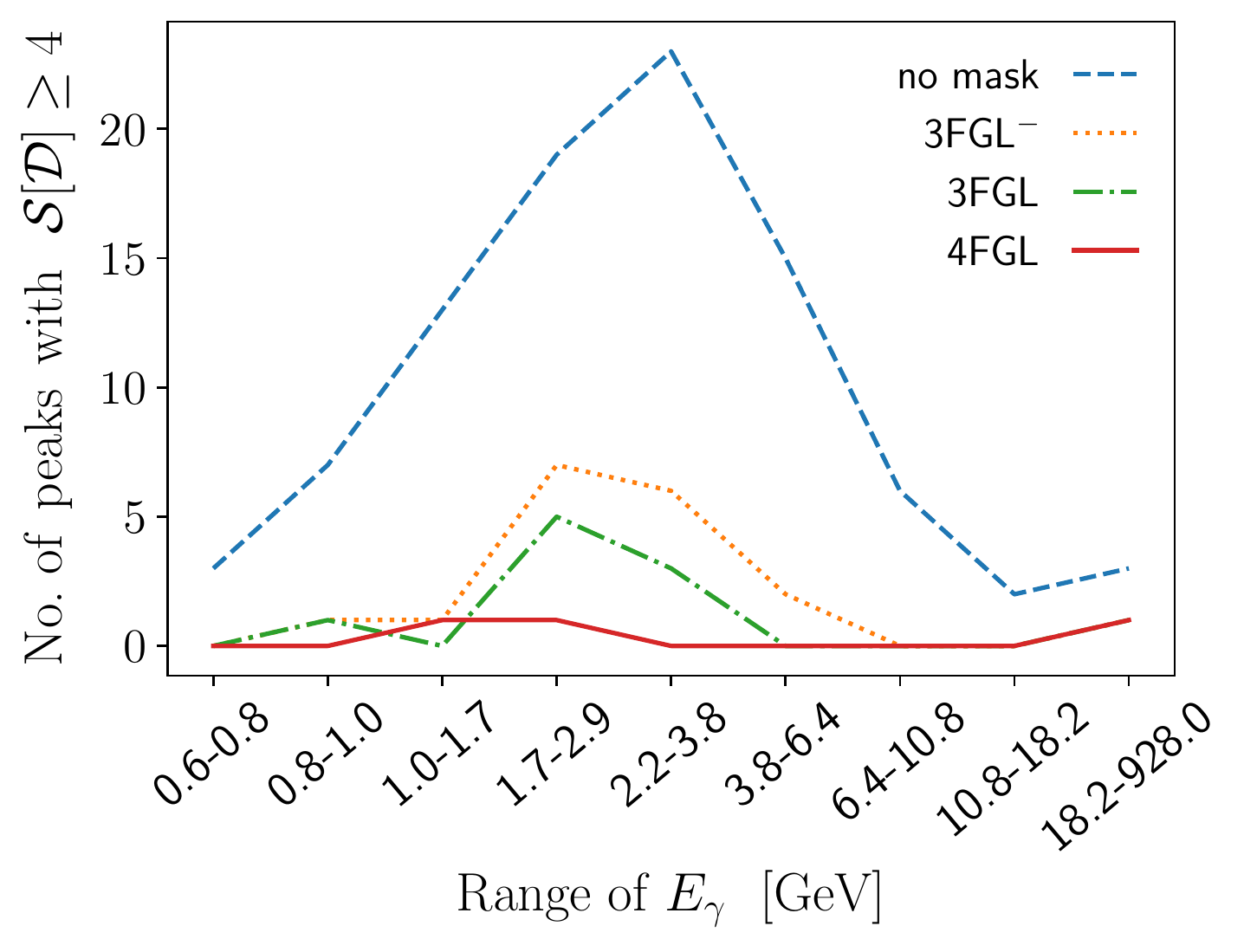}
\caption{Numbers of peaks over the entire sky with $\cS[\cD] \geq 4$ for $E_\gamma > 600\mev$, for two different energy binnings, and for four different choices of mask.}
\label{bin-compare}
\end{center}
\end{figure*}

We use Pass 8 data, version P8R3 recorded from Aug 4 2008 to Feb 20 2019 (weeks 9-559 of Fermi-LAT observations)\footnote{The Fermi-LAT data are publicly available at https://fermi.gsfc.nasa.gov/ssc/data/access/}. Using the Fermi {\tt ScienceTools P8v27h5b5c8} 
for selection event-cuts and to calculate the relevant exposure cube-files and exposure
maps\footnote{https://fermi.gsfc.nasa.gov/ssc/data/analysis/}, we select only the {\tt CLEAN}, FRONT-converted data, with the additional filter of {\tt zmax = $100^{\circ}$}. Our data maps are centered at the 
galactic center and cover a square window of $40^{\circ}$ side in galactic coordinates. Our pixels of size $0.1^{\circ} \times 0.1^{\circ}$  are on a Cartesian grid, thus do not have equal area which we account for. For the results in \Fig{mask-impact} we focus on the photons with energy 1.018 to 3.777 GeV (our 1 to 4 GeV range); see \Fig{bin-compare} for further energy dependence from the wavelet-based study.

\section{Wavelets at the Galactic Center}
The Mexican hat family of functions is defined by \cite{GonzalezNuevo:2006pa}
$M_n( \vec x, \sigma ) = \frac1{16\pi} \bigtriangleup^n e^{- |\vec x |^2/2\sigma^2}$, where $\bigtriangleup$ is the Laplacian operator and $\sigma$ is the width of the function. The second Mexican hat on the two-dimensional Cartesian grid is
\beq
 M_2( \vec x, \sigma ) = \frac{e^{ - |\vec x |^2/2\sigma^2 }}{16\pi \sigma^4}   \bL \pL \frac{|\vec x |^2}{\sigma^2 } - 4 \pR^2 - 8 \bR 
\label{mexhat2def}.
\eeq
The Mexican-hat matched filter applied to a map of photon counts $\cC$ is defined at a pixel $\Omega$ by
\beq \label{mexhatpix}
\cF[\cC|M_2](\sigma, \Omega) = \sum_{\Omega'} M_2 \!\pL ||\Omega-\Omega'||_p , \, \sigma \! \pR \cC[\Omega'],
\eeq
where $||\Omega-\Omega' ||_p$ represents the distance between the points $\Omega(') \equiv \{ b('), \ell(') \}$ in units of pixel  size $p$. We also make use of $\cF[\cC|(M_2)^2](\sigma, \Omega)$, obtained by squaring the kernel inside \Eq{mexhatpix} before performing the sum.

\begin{figure*}[t]
\begin{center}
\includegraphics[width=0.995\textwidth]{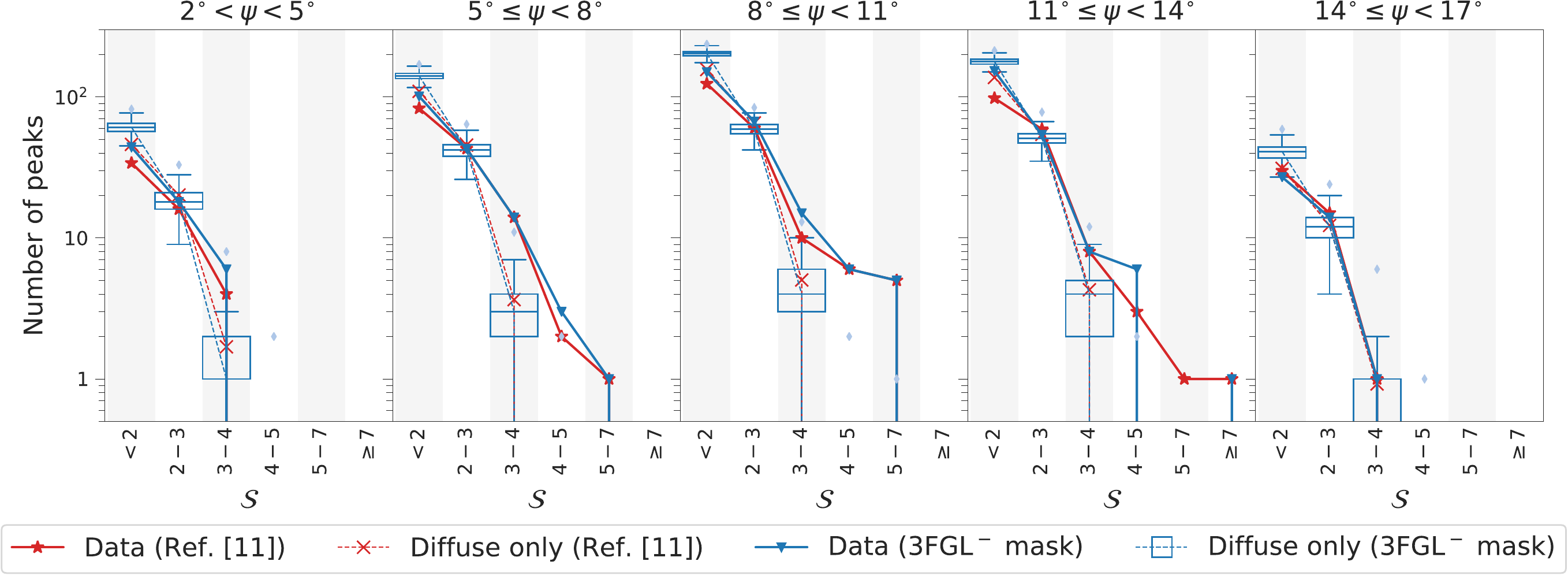}
\caption{Comparison to \cite{Bartels:2015aea}. We find excellent agreement in replicating these results for both $\cS[\cD]$ and $\cS[\cM]$.}
\label{compare-Bartels}
\end{center}
\end{figure*}

Following \cite{Bartels:2015aea}, we use the Mexican hat filter defined in \Eq{mexhatpix} with a width that varies with latitude. The intent is to approximately match the filter to the expected size of a point source observed by the Fermi-LAT instrument, as given by the instrument's point spread function (PSF). The choice for $\sigma$ in \cite{Bartels:2015aea} is 
\beq \label{width-Bartels}
\sigma = 0.4^\circ \times (0.53 + 0.3 |b|/12^\circ),
\eeq
where $0.4^\circ$ is approximately the Fermi-LAT PSF for photon energies $E_\gamma \sim \gev$, which has remained essentially unchanged between Pass 7 and Pass 8 Fermi-LAT data releases over the energies of interest for this study \cite{psf-vals}. We adopt \Eq{width-Bartels} in the bulk of this paper.

The test statistic used in \cite{Bartels:2015aea} is described as a signal-to-noise ratio,
\beq \label{SNR-M2}
\cS[\cC](\sigma, \Omega) \equiv \frac{\cF[\cC|M_2](\sigma, \Omega)}{\sqrt{\cF[\cC|(M_2)^2](\sigma, \Omega)}}~.
\eeq
A map of $\cS$ applied to Fermi data is shown in \Fig{S4-locations}, while the binned distribution of $\cS$ is presented in Figs.~\ref{mask-impact}, \ref{compare-Bartels}, \ref{dm-impact}, and \ref{csp-impact}.

Our clustering procedure is as follows. For each pixel on the map of $\cS$, we compare its $\cS$ value to those of its north, east, south, and west nearest neighbors, and keep the pixel if its $\cS$ value is the largest. Next, we perform the hierarchical clustering using \texttt{scipy.cluster.hierarchy} with the `euclidean' distance and `single' linkage criteria. This results in a hierarchical clustering dendrogram, which we then truncate at a maximum cophenetic distance of $0.3^\circ$ to get clusters of pixels.  We represent the $(b, \ell)$ and $\cS$ values  of each cluster of pixels with $(b, \ell)$ and $\cS$ values of the pixel with largest $\cS$ of that cluster. This procedure yields a map of peaks.

After obtaining the map of peaks, we mask: (1) the boundary region, which is $|\ell|, |b| \geq 12^\circ (20^\circ)$ for the analysis in \Fig{mask-impact} (Figs.~\ref{S4-locations} and \ref{spectra-compare}); (2) the disk region, $|b|< 2^\circ$; and (3) the neighborhood of point sources at $0.3^\circ$ around sources with 4FGL TS between 9 and 49 and $1.0^\circ$ around those with 4FGL TS greater than 49.
For the analysis of \Fig{mask-impact} we then subdivided the unmasked region into five sub-regions of $2^\circ-5^\circ$, $5^\circ-8^\circ$, $8^\circ-11^\circ$, $11^\circ-14^\circ$, and $14^\circ-17^\circ$ in projected angle from the Galactic Center. Within each sub-region, we binned the unmasked peaks according to their $\cS$ value into $(-\infty, 2)$, $[2, 3)$, $[3, 4)$, $[4,5)$, $[5, 7)$, and $[7, +\infty)$ bins.

For the analysis resulting in \Fig{spectra-compare}, we leave the 4FGL catalog unmasked and look for 4FGL sources within $0.55^\circ$ of each $\cS[\cD]\geq4$ pixel. This somewhat arbitrary definition allows us to associate ``peak 2'' of \Fig{S4-locations} to 
the source 4FGL J1750.0-3849, which is unassociated.
If the ``adjacent'' sources (by this definition) have no association or are of unknown origin, we add their spectra to produce the result in \Fig{spectra-compare}. If these sources are pulsars of unknown distance or are within 2 kpc of the Galactic center according to the ATNF pulsar catalogue \cite{1997ApJ...481..386B, 2016MNRAS.455.1751R}, we also add them. None of the sources that survive these cuts are classified as variable.

We also highlight an extended source near $\ell=-18^\circ, b=-2^\circ$. This source was also found and characterized in \cite{Balaji:2018rwz}. Because of its extension and its distance away from the disk, we remove it and all peaks within $1^\circ$ of this source from our $\cS[\cD]$ maps by hand when performing our analysis of \Fig{spectra-compare}.

\section{Further Comparison to Ref.~\cite{Bartels:2015aea}}

In \Fig{compare-Bartels} we show our attempt to replicate the results of \cite{Bartels:2015aea}. This information is not present in \Fig{mask-impact} because \cite{Bartels:2015aea} uses a different mask derived from the 3FGL catalog: we use the 3FGL catalog as a starting point for this mask, but omit 13 unassociated point sources with a hard spectrum and low variability, since these may in principle be bright but representative examples of a putative central point source population. This mask, which we refer to as ``3FGL${}^-$'', is identical to our 3FGL mask above, except with 13 fewer masked locations. Another difference in the analysis of \cite{Bartels:2015aea} is that they rely on a single diffuse emission model. They report results using only {\tt gll\textunderscore iem\textunderscore v06} model, which was the officially recommended model for searching for point sources at the time of publication.

The box-and-whiskers representing the spread of the models that we sampled, $\cS[\cM_i]$, encompasses the results of \cite{Bartels:2015aea} for the {\tt gll\textunderscore iem\textunderscore v06} model. We also check the {\tt gll\textunderscore iem\textunderscore v06} and find that it falls within the variation of our 60 models. Furthermore, we find very good agreement on the comparison of $\cS[\cD]$. Slight discrepancies are apparent: for instance, the $5 \leq \cS[\cD] < 7$ pixel they detect in $11^\circ \leq \psi < 14^\circ$ is apparently now found in our $4 \leq \cS[\cD] < 5$ bin. These slight discrepancies are likely due to the fact that we are using data which has been reprocessed since the publication of \cite{Bartels:2015aea}, and we are using front-converting Clean events instead of (all) Ultraclean events. However, gross statistics like the number of $\cS[\cD] \geq 4$ pixels are very compatible between the two analyses, lending further credence to the idea that the wavelet-based approach is indeed detecting point sources in a robust way.

Compared to the entire 3FGL mask, we see that the number of high-$\cS$ points increases slightly. This is due to the additional unassociated point sources that are allowed to remain unmasked when the 3FGL${}^-$ mask is employed.

\section{Energy-Dependence of Survival Function}

Because the statistics of $\cS$ are dependent on total counts, and because Poisson variation averages out in the large-$N$ limit while a true point source will only grow in significance, the number of pixels in any given $\cS[\cD]$ range can in principle depend on the energy range chosen for analysis. Here, we attempt to display the robustness of our results against choices of energy binning. We plot the ``survival function'', defined as the total number of pixels with $\cS[\cD] \geq 4$ across the $|\ell|\leq12^\circ, 2^\circ \leq |b|\leq12^\circ$ region of interest, for two choices of energy binning and for four choices of mask in \Fig{bin-compare}. Regardless of the choice of binning, we find that in every energy bin the number of point sources decreases to only very few pixels with $\cS[\cD] \geq 4$ upon applying the 4FGL mask. We have tested extensively and found the same effect with other binning choices. This indicates that the effect shown in \Fig{mask-impact} is present at all energies for which small-scale power is present and for all choices of energy binning.

\section{Effects of a Smooth Spatial GCE on $\cS$}

\begin{figure*}[t]
\begin{center}
\includegraphics[width=0.995\textwidth]{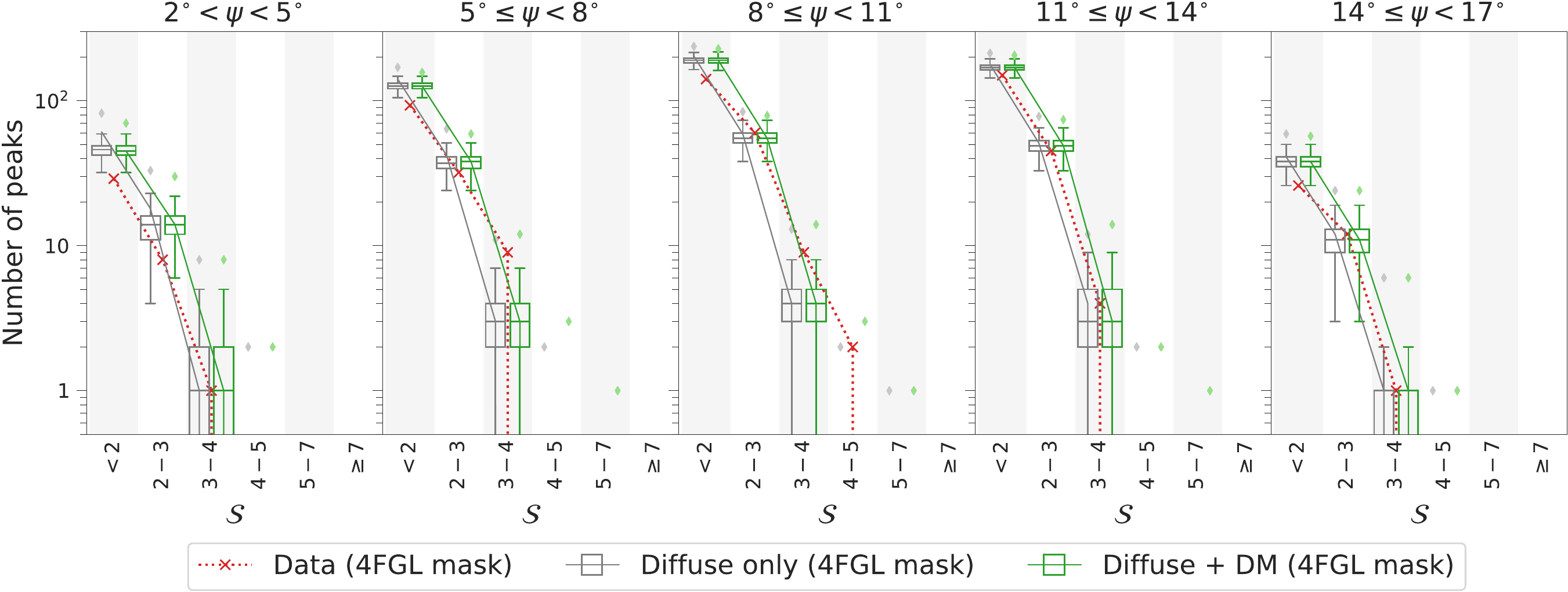}
\caption{Comparison of data to diffuse-only models and models that include diffuse emission plus a dark matter template. Because these overlap so well, we conclude that the wavelet-based search for small-scale power is not sensitive to new smooth emission.}
\label{dm-impact}
\end{center}
\end{figure*}

In \Fig{dm-impact}, we compare $\cS[\cD]$ against $\cS[\cM_i]$ for diffuse-only models and for models that include both diffuse emission and an entirely smooth spatial template. We model this smooth spatial template as arising from dark matter annihilation and normalize it to be bright enough to account for the entirety of the GCE. In the second box-and-whiskers of each $\cS$ bin, we combine the distributions for a squared NFW profile, a squared ``cuspy'' gNFW profile with $\gamma=1.2$, and a squared ``cored'' profile, which is a gNFW profile with an inner slope $\gamma = 0.8$. Despite the additional variance from this wider variety of models, we find that the diffuse-only expectation is very similar to the diffuse-plus-dark-matter annihilation in every spatial and significance bin.

Because these overlap so significantly, and the excess smooth emission does not produce any peaks with $\cS[\cD] \geq 4$, we conclude that the test of \Eq{SNR-M2} is not sensitive to new smooth emission. As a corollary, the lack of high-$\cS[\cD]$ peaks upon application of the 4FGL mask is not a contraindication of dark matter annihilation. We can calculate the negative log-likelihood for the number of pixels with $\cS[\cM_i] \geq 4$ for these dark matter simulations. We find $-2\ln\lambda(\bm \theta)|_{\rm DM} = 10.4$, which is a good fit for 15 bins of data.

\section{Effects of a Central Source Population on $\cS$}

\begin{figure*}[t]
\begin{center}
\includegraphics[width=0.995\textwidth]{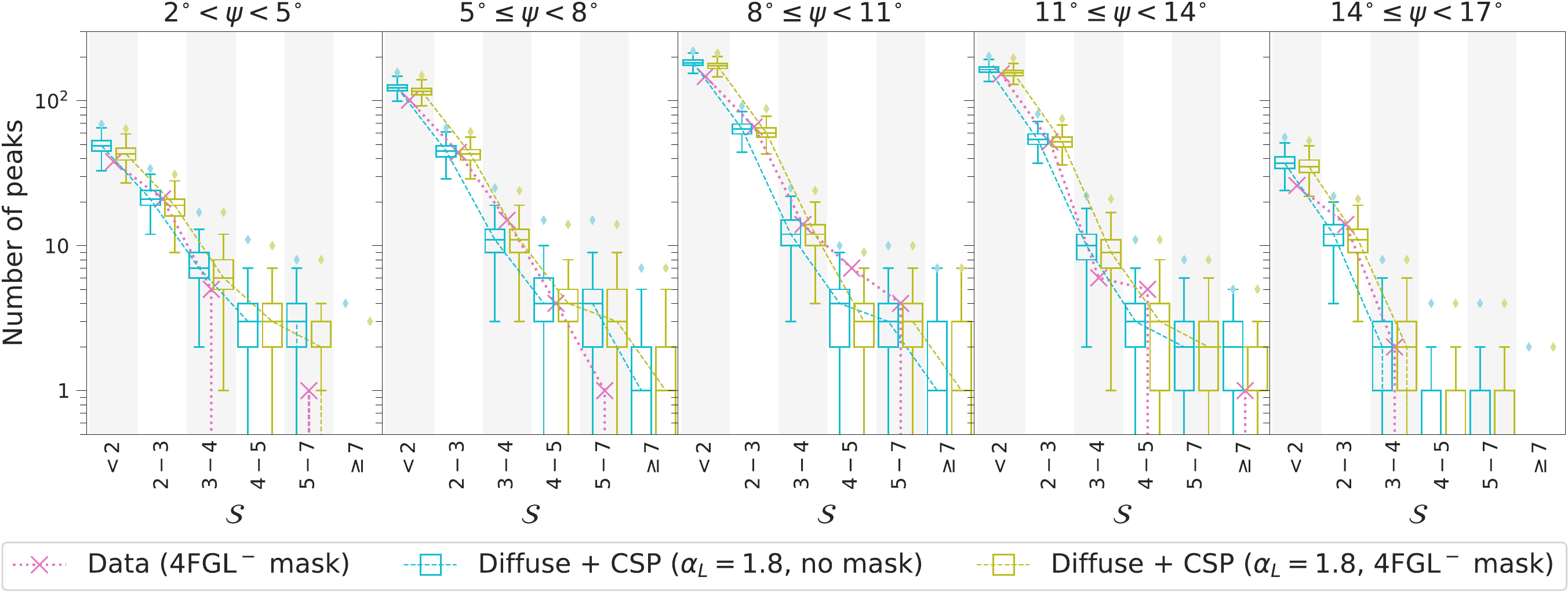}
\caption{Comparison of data to models that include diffuse emission plus a CSP. Because these are visibly discrepant, we conclude that the power law with index $\alpha_L=1.8$ produces too many bright sources, as discussed in the main text.}
\label{csp-impact}
\end{center}
\end{figure*}

If we interpret all unassociated and unknown 4FGL sources as possible bright representatives of a central source population (CSP) (which do not contribute to the template fit of \Fig{template_fit_result}, but which may nonetheless give us clues as to the nature of the CSP), we should construct a mask from the 4FGL catalog similar to the one described in the main text but omitting all unassociated and unknown sources. We refer to this mask as ``4FGL${}^-$.''

We compare the data with the 4FGL${}^-$ mask to simulations of CSPs that are bright enough to generate the GCE. Our simulations have hard cutoffs at $L_{\rm low} = 10^{29}\erg/\s, L_{\rm high} = 10^{35}\erg/\s$, and vary only by their power law indices $\alpha_L$, with $\alpha_L$ as defined in the main text. We model the CSP using ten distinct random realizations of CSP member locations, and we take ten independent Poisson samples of each resulting map; we use all 60 models as in our diffuse-only analysis in \Fig{mask-impact}. We show the results using either no mask or the 4FGL${}^-$ mask for $\alpha_L=1.8$ in \Fig{csp-impact}. In similar simulations for $\alpha_L =1.5 (2.0)$, we find more (fewer) pixels with $\cS[\cM_i] \geq 4$, especially in the inner $8^\circ$. We can compare the negative log-likelihood for the number of pixels with $\cS[\cM_i] \geq 4$ for these different CSP simulations. At 95\% CL, we expect $-2\ln\lambda(\bm \theta)<23.7$ for 14 degrees of freedom, while we find $-2\ln\lambda(\bm \theta)|_{\alpha_L=2.0} = 20.8$, $-2\ln\lambda(\bm \theta)|_{\alpha_L=1.8} = 27.1$, and $-2\ln\lambda(\bm \theta)|_{\alpha_L=1.5} = 30.6$. This supports our analytic results in the main text that if the GCE arises from a CSP, this population must have $\alpha_L \gtrsim 1.9$.

We also note that the insensitivity of our results to the mask and the relatively narrow spread despite the different spatial realizations are indications of the robustness of our procedure to finite-area effects from our choice of mask. Further tests of the CSP explanation of the GCE are warranted in light of our results.

\bibliography{w3}

\end{document}